\documentclass[article,nojss]{jss}

\usepackage{orcidlink,thumbpdf,lmodern}

\usepackage{amsmath,amssymb}

\newcommand{\copdCriticalTroughB}{-24.76}

\newcommand{\simonDesigns}{102}

\newcommand{\simonExampleError}{0.09313}
\newcommand{\simonExamplePower}{0.90284}
\newcommand{\simonExampleExpected}{14.546}

\newcommand{\studiolink}[2]{\href{https://ssm123ssm.github.io/statClaim/?open=#1}{#2}}

\author{Supun Manathunga~\orcidlink{0000-0002-2607-3430}\\McGill University}

\Plainauthor{Supun Manathunga}

\title{StatClaim: Study Designs as Types, Protocols as Programs and Statistical Claims as Proof Obligations}
\Plaintitle{StatClaim: Study Designs as Types, Protocols as Programs and Statistical Claims as Proof Obligations}
\Shorttitle{StatClaim}

\Abstract{
Statistical guarantees depend on the compatibility of a study's model,
hypotheses, sampling plan and analysis. StatClaim is a declarative
domain-specific language for encoding research protocols and statistical
analysis plans as programs built from objects whose types encode their
statistical roles within the study design.
The compiler checks type compatibility and applies contracts to check
additional requirements on the values and relationships of these objects.
When verification is requested, StatClaim's backend constructs and checks
proofs of supported statistical claims using \proglang{Lean} under stated
assumptions. These claims
concern Type~I error, familywise error, power and interval coverage, including
supported combinations of analyses and procedures with planned stopping
rules. Reports distinguish formally proved, conditional, cited and
author-declared guarantees and provide replayable certificates for verified
claims. The same specification supports sample-size searches, power
calculations and interval-precision assessment using exact calculations,
analytical approximations or reproducible simulation. A command-line
interface, editor integration and graphical block editor use a shared
compiler and execution layer. Validation reproduces 102 optimal and minimax
two-stage designs from Simon's published tables and verifies their error and
power bounds in \proglang{Lean}. A clinical trial case study illustrates
partial encoding of a published analysis plan using cited methods and
conditional proofs. The open-source implementation includes runnable
tutorials and examples for specifying and checking study designs before
outcomes are collected.
}

\Plainkeywords{study design, domain-specific languages, type systems,
formal verification, sample-size planning, Lean}

\Address{
  Supun Manathunga\\
  Division of Experimental Medicine\\
  McGill University\\
  Montreal, QC\\
  Canada\\
  E-mail: \email{supun.manathunga@mail.mcgill.ca}
}

\begin{document}

\section[Introduction]{Introduction} \label{sec:intro}

A statistical analysis can be calculated correctly while its use in a study is not justified. Certain properties of a test, such as its Type~I error guarantee, depend on other elements such as the null hypothesis, sampling mechanism and decision rule to which it applies.
Testing several endpoints separately at the 5\% level, for example, does not by
itself establish a 5\% familywise error bound. Likewise, a guarantee for a single
analysis at a fixed sample size does not automatically extend to a procedure
that repeatedly examines accumulating data and stops at significance. Identifying these questions and addressing them during protocol development allows researchers to revise the
design before data collection begins.

StatClaim is a domain-specific language (DSL) for expressing the
statistical components of a research protocol as typed objects, including
models, sampling plans, hypotheses, analyses and claims \citep{statclaim}.
Domain-specific types distinguish these concepts and declarations
construct values whose relationships are checked by compatibility contracts.
For example, an analysis must support the declared null and use the model to
which that null belongs. The compiler reports incompatible declarations and
unresolved requirements without executing statistical calculations. Compilation alone does not
establish a statistical guarantee. When verification is requested, StatClaim's
backend constructs and checks proofs of supported statistical claims using
\proglang{Lean} \citep{deMoura+Ullrich:2021}.

Executable design specification and formal checking of statistical programs
have important precedents. \pkg{DeclareDesign} represents research designs
through models, inquiries, data strategies and answer strategies, and diagnoses
properties such as power, bias and coverage by simulation
\citep{Blair+Cooper+Coppock:2019,DeclareDesign}.
\pkg{StatWhy} verifies annotated hypothesis-testing programs through the Why3
platform, checking requirements for the use of statistical methods
\citep{Kawamoto+Kobayashi+Suenaga:2025}. StatClaim combines a language focused on
study design with explicit compatibility contracts, graded evidence and, for
supported methods, formal proofs of the statistical guarantees themselves.

StatClaim focuses primarily on the design phase, with the goal of making a protocol's statistical justification
explicit, inspectable and, where supported, formally verifiable before outcomes
are observed. Claims concern Type~I error, familywise error, power or interval
coverage. Composition supports procedures involving multiple endpoints or
planned interim analyses, with requirements for the whole declared procedure.
An unresolved requirement indicates missing justification, but it need not imply that the proposed statistical claim is false. Successful checks explain why compatibility contracts hold and report the
assumptions on which claims depend. These explanations are separate from proofs
of statistical guarantees. Evidence grades distinguish guarantees proved in
\proglang{Lean}, compositions proved conditional on assumed member guarantees,
results cited from the literature and author-declared statements.

The same specification supports study planning through power assessment,
evaluation of interval precision and sample-size searches. Depending on the
method, these use exact calculations, analytical approximations or reproducible simulation with reported Monte Carlo uncertainty. Supported exact planning
claims can certify that a design meets its error and power targets and, where
available, that no smaller candidate in a declared search range is sufficient.
Power is assessed at a stated planning alternative. Precision calculations and
simulation estimates currently provide numerical assessments without
\proglang{Lean}-checked guarantees.

Compilation makes study reports available for review. Successful formal
verification produces replayable proof certificates. Researchers can write a
protocol as a \code{.stat} source file using a text editor with language hints
from the editor integration, or use StatClaim Studio, a graphical block editor
backed by the same compiler and statistical contracts.

As a simple illustrative example, the following program declares two independent binomial samples
with fixed arm sizes, specifies an equal-proportions null hypothesis and requests
verification of the two-sided Fisher test's Type~I error bound. No observed data
are required and the claim concerns the procedure's repeated-sampling behavior for
every common success probability under the null.
\begin{CodeInput}[samepage=true]
use models.two_arm_binomial
use sampling
use tests.fisher

let study = two_arm_binomial(n_control=50, n_treatment=50)
let design = fixed(model=study)
let null = equal_proportions(model=study)
let analysis = fisher(design=design, null=null, alpha=1/20)

let claim = typeI(analysis)
let verification = verify(claim)
\end{CodeInput}
\studiolink{paper/fisher_design}{Open this example in StatClaim Studio.}
Execution environments are described in Section~\ref{sec:studio}.
Running \code{statclaim check study.stat} reports the established guarantee and
its assumptions. The following output is excerpted and line-wrapped:
\begin{CodeOutput}
verified [verification]: typeI(analysis)
  Evidence: proved in Lean; guarantee: finite-sample bound
  The specified two-sided Fisher analysis has Type I error at most
  1/20 under the null, for every common success probability in [0, 1].
  Assumptions:
    The samples follow two independent binomial distributions.
    Positive arm sizes were fixed before outcomes were observed.
    The error bound is under the null of equal success probabilities,
    for every p in [0, 1].
Program status: verified
\end{CodeOutput}
Changing only the sampling plan to unspecified data-dependent stopping,
\begin{CodeInput}
let design = adaptive(model=study)
\end{CodeInput}
leaves the guarantee unresolved. The checker reports the following diagnostic
(source location omitted and lines wrapped):
\begin{CodeOutput}
obligation [sampling_scope]: analysis: data-dependent stopping needs
a justified sampling law before the fixed-sample guarantee can be used.
This is not evidence of inflated error.
\end{CodeOutput}
The diagnostic identifies missing justification. A guarantee for data-dependent
stopping requires a specification of the complete procedure and a suitable
theorem. The complete variant can also be
\studiolink{paper/fisher_adaptive}{opened in StatClaim Studio}.

The current library includes exact methods for binomial and selected
observational designs, randomization tests and two-stage single-arm designs,
alongside contracts for cited methods such as t-tests, log-rank tests and
regression analyses.

The contribution of StatClaim is the integration of typed study specifications,
explainable compatibility checks, planning operations and explicit evidence for
statistical claims. Section~\ref{sec:models} introduces the statistical objects,
claims and compatibility contracts. Section~\ref{sec:illustrations} describes
the language and compiler architecture. Section~\ref{sec:verification} explains
contract derivations, formal verification and certificate replay.
Section~\ref{sec:planning} covers power, sample-size and precision planning,
including simulation-based assessment. Section~\ref{sec:interfaces} presents
the authoring interfaces and method-extension workflow.
Section~\ref{sec:validation} checks published two-stage designs and presents
the COPD analysis-plan example. Section~\ref{sec:summary} discusses scope,
limitations and further development. Appendix~\ref{app:tutorials} summarizes
the tutorial and example coverage.

\section{Statistical claims and contracts} \label{sec:models}
\suppressfloats[t]

\subsection{Objects and their statistical meaning}

StatClaim separates the objects needed to describe a study from the evidence
needed to justify a claim about it. Table~\ref{tab:objects} summarizes the main
roles. A model specifies a family of probability laws and an admissible outcome
domain. A sampling plan describes how observations are acquired, including the
stopping mechanism. An analysis binds a design, a null hypothesis and a
procedure with its settings. For cited methods, model declarations and listed
assumptions need not supply a fully formalized probability law.

\begin{table}[htbp]
\centering
\begin{tabular}{p{0.22\textwidth}p{0.70\textwidth}}
\hline
Object & Role \\
\hline
Model & Parameter domain, possible observations and sampling assumptions. \\
Sampling plan & How sampling proceeds and when it ends. \\
Hypothesis & A restriction under which a guarantee is evaluated. \\
Analysis & A design, supported hypothesis, procedure and settings. \\
Observation & Supplied data checked against the declared domain. \\
Claim & A specified property of an analysis, result or planning target. \\
Execution result & The outcome of a calculation, simulation or verification request. \\
Evidence & The basis for an established statement, with assumptions and provenance. \\
\hline
\end{tabular}
\caption{Principal object roles. A particular declaration constructs a value, and its type distinguishes the role it can play in subsequent declarations.}
\label{tab:objects}
\end{table}

For a fixed-sample model, let $\Theta$ be the parameter space and $P_\theta$ a
probability law on an outcome space $\mathcal{X}$. An analysis specifies a null
set $H\subseteq\Theta$ and a measurable rejection event $R\subseteq\mathcal{X}$.
Its Type~I error claim at level $\alpha$ is
\begin{equation}
  \forall\theta\in H,\qquad P_\theta(R)\leq\alpha.
  \label{eq:typeI}
\end{equation}
The sampling plan matters because the law attributed to its observations must
agree with the law used by the theorem. In the shared \proglang{Lean} core,
\code{Compatibility} explicitly supplies this agreement. A fixed plan can be
constructed with the model's law, but an unspecified adaptive plan does not inherit
that law. This leaves an open requirement for a justified sampling law.

For interval estimation, let $\tau(\theta)$ denote the quantity of interest,
such as a response rate or a difference between rates. Let $X$ denote random
observations with law $P_\theta$, and let $C(x)$ be the interval returned by the
procedure for possible observations $x$. A coverage claim at confidence level
$1-\alpha$ states that
\begin{equation}
 \forall\theta\in\Theta,\qquad
 P_\theta\{\tau(\theta)\in C(X)\}\geq1-\alpha.
 \label{eq:coverage}
\end{equation}
For each fixed parameter value, the interval includes the quantity of interest
with probability at least $1-\alpha$ over repeated samples from the model.

When sampling can stop early, the recorded observations also depend on the
stopping rule. For example, a two-stage binomial study first observes a response
count $X_1$ among $n_1$ participants. It stops for futility if $X_1$ is at most
a prespecified threshold $r_1$. Otherwise, it observes a second count $X_2$
among $n_2$ additional participants and applies the final rejection rule. The
record contains either the first count and an early-stop indicator, or both
counts and a completion indicator.

The formal core represents this mechanism as a sampling policy. Let $Z$ denote
the potential data if all planned stages were observed, with possible values
in a space $\mathcal{Z}$ and probability law $L_\theta$. In the two-stage
example, $Z=(X_1,X_2)$ has the declared product-binomial law. A measurable rule
$g\colon\mathcal{Z}\to\mathcal{Y}$ maps the potential data to the record
$Y=g(Z)$, whose possible values lie in $\mathcal{Y}$. For any measurable set
$B\subseteq\mathcal{Y}$ of records, its probability is
\begin{equation}
 Q_\theta(B)=L_\theta\{z\in\mathcal{Z}\mid g(z)\in B\}.
 \label{eq:record-law}
\end{equation}
Thus $Q_\theta$ is the distribution of the recorded data after applying the
stopping rule. Type~I error and power are evaluated using this distribution
and a rejection rule defined on the records. The implemented two-stage module
specifies these components.

StatClaim supports explicit stopping rules through complete design
constructors, including two-stage binomial designs. The generic
\code{adaptive(model=study)} declaration instead marks an unspecified
data-dependent mechanism. A supported guarantee requires the mechanism to be
specified through an appropriate design constructor.

\subsection{Compatibility, assumptions and open requirements}

Type-compatible declarations undergo additional checks on their values and
relationships. This has a conceptual connection to refinement types, which
express more precise restrictions within an existing type
\citep{Freeman+Pfenning:1991,Rondon+Kawaguchi+Jhala:2008}. For example, an
analysis constructor requires a hypothesis belonging to the design's model and
supported by the chosen method. StatClaim implements these restrictions through
registered compatibility contracts evaluated during elaboration. It does not
currently provide general refinement-type inference or predicate-based
subtyping.

A test for binomial counts cannot consume a continuous-outcome model. For the two-sided Fisher
analysis, an equality of proportions is supported, whereas a nonzero equality
of differences is a different statistical question. The exact-zero difference
is normalized to equal proportions using an equivalence also established in
the formal library. The checker does not attempt arbitrary equivalence proofs between hypotheses.

StatClaim rejects declarations that combine incompatible objects. If the declarations
are valid but the requested guarantee lacks the required justification, the
claim remains unresolved.
Researchers must assess whether the assumptions surfaced for valid programs hold in the actual study.

Design declarations can expose contradictions with a method's
assumptions. Declaring several measurements per participant prevents treating
those measurements as independent participants with an ordinary two-sample
t-test. Declaring cluster assignment prevents substituting participant counts
for the number of independent clusters. Outcome direction can also be stated
explicitly and checked against a one-sided null. If these facts are omitted,
the relevant assumptions remain visible.

\subsection{Evidence and scope}

Table~\ref{tab:evidence} distinguishes the basis of a statement from the status
of its execution. A cited method can pass its compatibility checks without a
formal proof of its statistical guarantee. A conditional result establishes a
composition theorem whose member bounds remain premises. A declared method
records an author's assertion and checks the supported design requirements.
It supplies no mathematical justification for that assertion.

\begin{table}[t]
\centering
\begin{tabular}{p{0.18\textwidth}p{0.16\textwidth}p{0.55\textwidth}}
\hline
Evidence grade & Result status & What establishes the statement \\
\hline
Proved & \code{verified} & A proof checked by \proglang{Lean}. \\
Conditional & \code{conditional} & A checked combination, assuming its member guarantees. \\
Cited & \code{cited} & A literature result whose supported contract fits the declaration. \\
Declared & \code{declared} & An author assertion with no checked guarantee. \\
\hline
\end{tabular}
\caption{Evidence grades of statements in StatClaim.}
\label{tab:evidence}
\end{table}

The scope of evidence is recorded separately. A finite-sample guarantee applies at the declared
sample size, while an asymptotic statement describes limiting behaviour. A
conditional combination can inherit an asymptotic member's limitation. Each
result identifies its weakest member evidence, and the file summary reports
the weakest part of its evidence-bearing results.

\subsection{Composition of procedures}

Composition combines the guarantees of individual analyses into a guarantee
for a procedure involving several tests. For example, a study may test two
endpoints and require the probability of any false rejection to be at most
5\%. A Bonferroni procedure can allocate 2.5\% to each test. If each test
controls its own Type~I error at that level, the probability of rejecting any
true null is at most the sum of the allocated levels, or 5\%. The endpoints
need not be independent.

More generally, consider $m$ tests. For test $i$, let $H_i$ be its null
hypothesis, $R_i$ the event that it rejects, and $\alpha_i$ its error bound when
that null is true. At a parameter value $\theta$, write $I_0(\theta)$ for the
indices of the true null hypotheses. Familywise error is the probability of
rejecting at least one of those hypotheses. The union bound gives
\begin{equation}
 P_\theta\!\left(\bigcup_{i\in I_0(\theta)}R_i\right)
 \leq \sum_{i\in I_0(\theta)}\alpha_i
 \leq \sum_{i=1}^{m}\alpha_i\leq\alpha.
 \label{eq:bonferroni}
\end{equation}
The bound holds for every combination of true and false nulls, provided the
individual error bounds hold. StatClaim checks that the declared levels sum to
at most the requested family level $\alpha$. The proved Fisher-family adapter
also establishes the individual bounds under its stated sampling assumptions.
The general Bonferroni adapter uses \proglang{Lean} to check the combination
while the individual bounds remain assumptions, giving conditional evidence.

Fixed-sequence testing uses a different rule. Endpoints are tested in a
prespecified order, each at the full level $\alpha$, and testing stops when an
endpoint is not rejected. Any false rejection requires rejection of the first
true null in that order, whose own error probability is at most $\alpha$.
\proglang{Lean} checks this argument conditional on the individual test bounds.

Planned interim analyses repeat one question at several declared sample sizes.
Each look receives part of the error budget, and the study stops at the first
rejection. StatClaim checks that the allocated levels sum to at most $\alpha$
and applies the union bound across looks. The resulting claim concerns Type~I
error for that one question. The current construction assumes each look's
bound and does not use correlation between looks to obtain less conservative
thresholds.

Graphical procedures allow a rejected endpoint's allocated level to be passed
to other endpoints according to a declared graph. StatClaim checks the
starting allocations and transfer rules, while the procedure's general
guarantee is cited. A family or sequence can also contain a whole supported
procedure as one entry. The resulting evidence retains the assumptions and
grades of the parts on which it depends.

\section{Language and architecture} \label{sec:illustrations}

\subsection{Declarations and exact inputs}

A \code{.stat} file is a sequence of module imports, immutable bindings and
explicit operation requests. The following schematic grammar summarizes its
declaration structure. Constructor signatures supply the domain-specific
argument constraints.
\begin{Code}[samepage=true]
statement  = "use" module
           | "let" name "=" expression
           | operation
expression = literal | name | call | list | expression "." name
operation  = verify(claim)
           | calculate(claim)
           | simulate(claim, seed=integer, ...)
\end{Code}
The ellipsis denotes optional simulation settings. Imports
select constructors from the installed registry. They do not load arbitrary
\proglang{Python} code or other \code{.stat} files. Names must be defined before
use and cannot be rebound. Calls allow positional arguments followed by named
arguments. Comments begin with \code{\#}.
The current language version (v0.41.0) has no general arithmetic expressions, loops, conditionals or
embedded host-language evaluation. More elaborate statistical procedures are
provided through registered constructors with explicit semantics.

Integers, finite decimals and fractions are exact inputs. For example,
\code{0.025} and \code{1/40} denote the same rational value. A fraction literal
does not permit division of arbitrary expressions. The frontend checks domains
such as positive sample sizes, probability levels and count bounds, with
method-specific restrictions where needed. Exact input representation does not
mean that every backend calculation is exact. Analytical power approximations
and simulation use numerical algorithms whose status is reported separately.

Models, hypotheses and analyses have distinct roles,
and constructors require the appropriate input categories. A particular study
is a value of such a category. Its arm sizes and stopping rule also participate
in semantic checks. The frontend does not encode every statistical condition as
a separate dependent type. The \proglang{Lean} backend uses model- and
claim-indexed objects for mathematical statements and evidence.

\subsection{Composing models and naming claims}

Two-arm models can be constructed from an assignment and an outcome, allowing
several outcomes to share the same allocation. The following self-contained
example declares continuous scores for 100 participants per arm:
\begin{CodeInput}[samepage=true]
use models.two_arm_continuous

let arms = assignment(n_control=100, n_treatment=100)
let score = continuous_outcome(better="lower")
let study = two_arm(assignment=arms, outcome=score)
\end{CodeInput}
\studiolink{paper/composed_model}{Open this example in StatClaim Studio.}
Sharing an assignment reuses its declared structure without asserting
independence between outcomes. Additional fields can describe randomization,
clustering, repeated measurements or analysis population. Outcomes may be named
so that repeated testing of the same declared question can be detected within
a family. The meaning of those names must still be supplied accurately.

Claims can be named and reused. In the introductory program,
\code{typeI(analysis)} constructs a claim before verification is requested. It can
therefore be inspected or reused before any backend runs. Observed-result
claims are constructed separately. Continuing the introductory example,
\begin{CodeInput}[samepage=true]
let data = observe(analysis=analysis,
                   control_successes=12, treatment_successes=24)
let observed = result(data)
let computation = calculate(observed)
let checked_result = verify(observed)
let saved = checked_result
let p_value = computation.calculation.p_value
let certificate = checked_result.evidence.certificate.file
\end{CodeInput}
\studiolink{paper/fisher_observed}{Open the complete program in StatClaim Studio.}
Here \code{calculate} computes a proposed result and decision, while
\code{verify} requests independent formal checking of the supported result
properties. Aliasing \code{checked\_result} does not run verification again.
Neither the observed counts nor their p-value establish the truth of the null.

\subsection{Public fields and execution phases}

Only fields in the public schema are accessible. An analysis exposes its
design, null and level. A claim exposes its subject and scope. Execution results
expose statuses, diagnostics and the appropriate result payload. The same
schema drives field checking, completion, hover information and serialization.
Field access does not provide arbitrary access to the underlying
\proglang{Python} attributes or methods.

Some projections are known while compiling. For example,
\code{analysis.alpha} can configure another compatible constructor. Others
depend on execution. The compiler retains a typed deferred projection for
\code{computation.calculation.p\_value}, which is resolved during execution. Such a value
can be named, aliased or collected in a list, but it cannot configure a new
statistical constructor. That constructor's inputs must be known during
compilation. This prevents a deferred result from silently changing the study
whose claims are being inspected.

Evidence has an optional type. Its projection is pending before execution and
absent after an unsuccessful verification attempt. These are different states.
Access through an absent optional value
propagates absence, so a certificate path cannot be treated as present merely
because its field expression was well typed. Cited evidence can be present
without a formal certificate. Optional values are retained in the result
schema, and they are not implicitly converted into proof objects.

\subsection{Compiler and execution pipeline}

Figure~\ref{fig:pipeline} shows the shared architecture. The parser creates a
syntax tree with source spans. Elaboration resolves imports and names, checks
argument categories and constructor contracts, and builds immutable typed
objects. Requirement inspection adds known obligations for the requested
claims. Bindings retain argument locations and named dependencies so that a
diagnostic can identify both the failing use and the declaration it refers to.

\begin{figure}[t]
\centering
\begin{tabular}{cc}
\multicolumn{2}{c}{\code{.stat} source} \\
\multicolumn{2}{c}{$\downarrow$} \\
\multicolumn{2}{c}{Parsing and typed elaboration} \\
\multicolumn{2}{c}{$\downarrow$} \\
\multicolumn{2}{c}{Contracts, requirements and execution plan} \\
$\swarrow$ & $\searrow$ \\
Editor diagnostics and inspection & Explicit operation requests \\
 & $\downarrow$ \\
 & Calculation, simulation or verification \\
 & $\downarrow$ \\
 & Results, evidence and retained source snapshot
\end{tabular}
\caption{Compilation is shared by the interfaces and does not execute the
requested operations. Formal verification is one explicit execution route.}
\label{fig:pipeline}
\end{figure}

The compiler is implemented in \proglang{Python}. It constructs objects in
memory.
\code{statclaim compile study.stat} performs the fast, backend-free checks.
\code{statclaim check study.stat} compiles the entire file and then executes
eligible requests in declaration order. An invalid file does not start the statistical
backend. A valid file can contain an unresolved request alongside other
requests that can execute, and the report records the status of each and the
overall unresolved condition. No Lean proof, numerical calculation or
certificate write is required to display compilation diagnostics.

Execution resolves the typed result slots using request identifiers. The run
retains the source, normalized objects and result report. Additionally, formal requests can
add certificate bundles. If a source file changes during execution, results
refer to the retained snapshot and the current file is marked stale. This
binding between source and result is essential because modifying an input can
change both the calculation and the proposition requiring proof.

\section{Contract derivation and formal verification} \label{sec:verification}

\subsection{Explaining and checking a contract derivation}

For supported declarations, the compiler retains a derivation explaining why
the contract holds. A derivation records the subject, versioned rules, facts
used by each rule, dependencies between steps, assumptions and open
obligations. For a two-sided Fisher analysis, the compiler checks that the
design and null refer to the same model, that the method supports the declared
equality-of-proportions null and that the level lies in its permitted range.
The derivation also records that the declared sampling plan is fixed, as
required for the Fisher Type~I error guarantee. When verification is requested,
the backend applies the corresponding theorem to the declared design and
level, and \proglang{Lean} checks the resulting proof.

The command \code{statclaim explain study.stat} presents these steps on
request. An independent checker re-evaluates the premises from the serialized
subject and checks the derivation's structure, conclusions, assumptions and
open requirements. It shares the rule statements with the derivation builder
but does not call the builder or the constructor's contract-checking functions.
This provides a separate check of the recorded explanation. An altered rule
name or successful-status flag is insufficient. It is not a formal proof of
the frontend or of the statistical correctness of the rule definitions.

\code{statclaim report study.stat} aggregates claim descriptions, evidence
expectations and an assumption register suitable for inspecting a statistical
analysis plan. Producing this report does not execute verification. It can
describe what a successful request would establish while the request is still
pending. \code{statclaim impact before.stat after.stat} compares declarations
and their dependencies to identify affected contracts and claims requiring
rechecking.

\subsection{Formal objects and theorem application}

The shared formal library uses \proglang{Lean} and \pkg{mathlib}
\citep{deMoura+Ullrich:2021,mathlib2020}. Its model object contains a probability
law for every parameter, an admissible observation set and evidence of
measurability and support. An analysis carries its design, hypothesis and
procedure. A claim contains a proposition, and a logical certificate contains
proof of that proposition. These logical certificates are distinct from the
serialized artifact bundles described below. A label or an assumed assessment
cannot be projected into a proof.

For the introductory example, the reusable Fisher theorem bounds the rejection
probability of a two-sided test using probability ordering by
$\alpha$ under two independent binomial samples with a common success
probability. The theorem quantifies over sample sizes and every real
$p\in[0,1]$. Its proof first
establishes the finite-distribution p-value bound, constructs the normalized
hypergeometric probabilities and factors the joint binomial probabilities by
the pooled total. Summing the conditional bounds gives the unconditional
bound. The formal connection to \pkg{mathlib}'s probability measures makes this
a probability statement about the declared model. The argument also covers
$p=0$ and $p=1$.

The Fisher provider requires sampling compatibility and instantiates that
theorem for the declared design and level. A design-level request therefore
needs no observations. Other modules supply different statements. One-sided
tests and two-stage designs can use monotonicity to extend checked boundary
calculations over a composite null, while interval modules establish coverage.
A new constructor does not acquire these theorems merely by returning an
object with a similar interface.

\subsection{Checking a concrete calculation}

\begin{figure}[!htb]
\centering
\begin{tikzpicture}[
  font=\small,
  box/.style={draw=black!55, rounded corners=2pt, align=center,
    inner sep=7pt, line width=0.5pt},
  flow/.style={->, >=stealth, draw=black!65, line width=0.6pt}
]
\node[align=center] (inputs) at (0,0) {
  \textbf{Arm sizes and observed counts}\\[3pt]
  $n_C,\ n_T,\ x_C,\ x_T$
};
\node[box, text width=5.9cm, minimum height=1.8cm] (python) at (-3.6,-1.9) {
  \textbf{\proglang{Python}}\\[4pt]
  Integer-weight recurrence\\[4pt]
  Proposed p-value $v=a/b$
};
\node[box, text width=5.9cm, minimum height=1.8cm] (lean) at (3.6,-1.9) {
  \textbf{\proglang{Lean}}\\[4pt]
  Proved factorial formula\\[4pt]
  $T(x)=A/B$
};
\draw[flow] (inputs.south) -- (0,-0.7) -| (python.north);
\draw[flow] (0,-0.7) -| (lean.north);
\node[box, text width=12.9cm, minimum height=1.3cm] (check) at (0,-4.4) {
  \textbf{Natural-number proof obligation}\\[4pt]
  $Ab=aB$\\[4pt]
  Checked by \code{decide +kernel} through \proglang{Lean}'s kernel
};
\draw[flow] (python.south) -- (-3.6,-3.1) -- (0,-3.1) -- (check.north);
\draw[draw=black!65, line width=0.6pt]
  (lean.south) -- (3.6,-3.1) -- (0,-3.1);
\node[box, text width=12.9cm, minimum height=1cm] (result) at (0,-6.4) {
  \textbf{Verified p-value for the supplied counts}\\[4pt]
  $v=T(x)$
};
\draw[flow] (check.south) -- (result.north);
\end{tikzpicture}
\caption{Verification of a Fisher p-value for supplied counts. The arm sizes
are $n_C,n_T$ and the observed success counts are $x_C,x_T$. The numerators
$a,A$ and denominators $b,B$ are natural numbers with $b,B>0$. The certificate
also checks the count bounds and rejection decision at the declared level
$\alpha$. The Type~I error bound follows from a separate procedure theorem.}
\label{fig:fisher-result-check}
\end{figure}
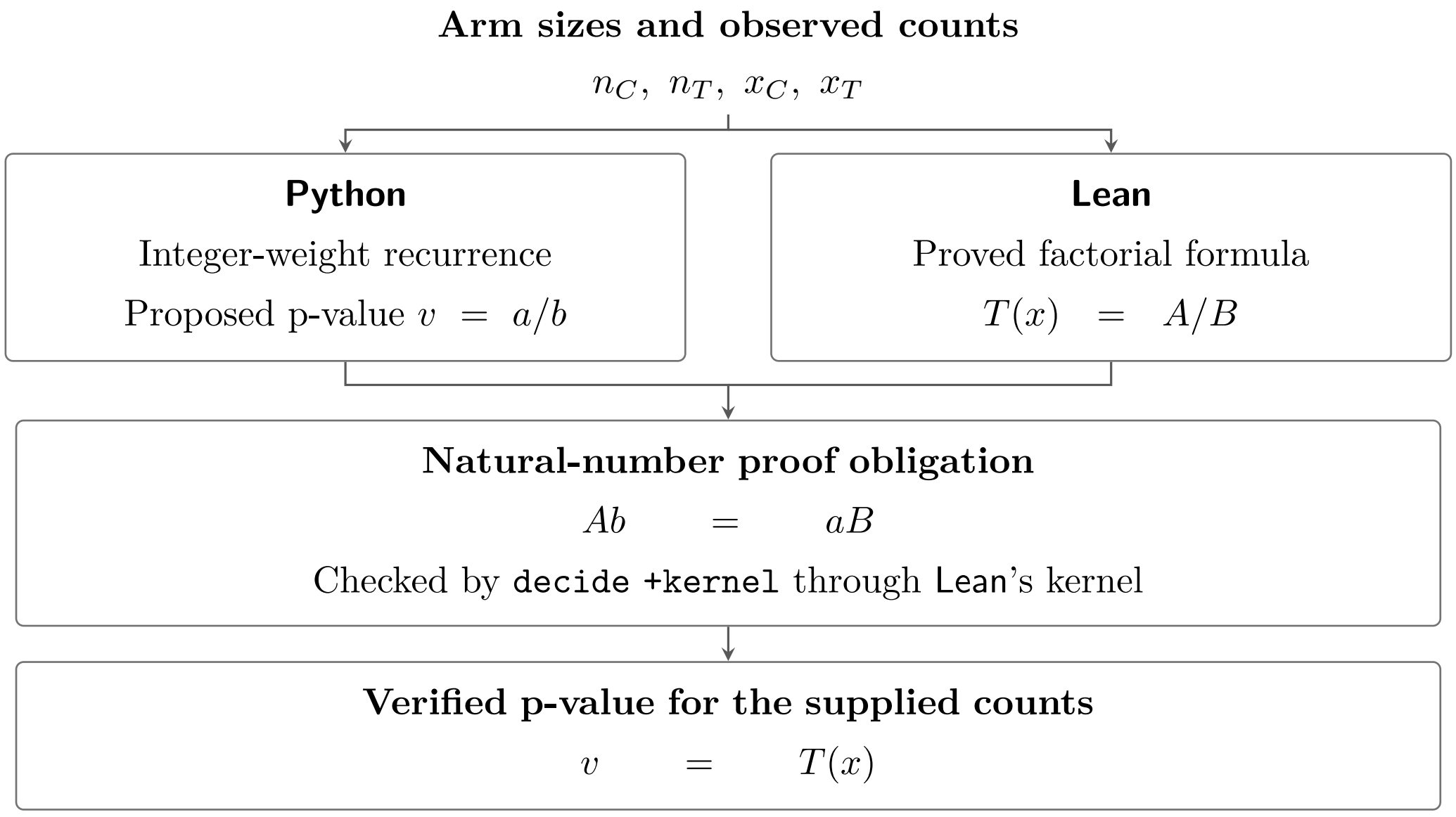

The procedure-level theorem and correctness of an observed calculation are
separate propositions. If $T$ is the formal procedure, $x$ is an admissible
observation and $v$ is a proposed output, the latter claim is
\begin{equation}
 v=T(x).
 \label{eq:result}
\end{equation}
For Fisher's test, \proglang{Python} proposes an exact rational p-value using
an integer-weight recurrence. The formal library supplies a factorial formula
proved equal to the mathematical definition. Figure~\ref{fig:fisher-result-check}
shows how these two expressions are connected. With positive denominators,
equality of the fractions reduces to a natural-number equality, checked by
\code{decide +kernel} through \proglang{Lean}'s kernel.

This checks a particular calculation against the specified mathematical
procedure. It does not prove that every execution of the \proglang{Python}
implementation is correct, or that the supplied counts arose through the
declared sampling mechanism. Conversely, a numerical calculation alone is not
evidence for the uniform error claim in Equation~\ref{eq:typeI}.

\subsection{Certificate issuance and replay}

Before issuing a formal certificate, the backend builds the project, checks
the running \proglang{Lean} version against the pinned toolchain and checks
dependency revisions. It binds the request to canonical inputs and records
hashes of relevant application and formal sources. After running the generated
proof, it checks that the bound sources and artifacts remained stable.
Unexpected dependency changes, proof failure or a timeout leave the claim
unresolved and do not issue a certificate.

Each required theorem is subjected to a transitive axiom audit. The permitted
logical dependencies are \code{propext}, \code{Classical.choice} and
\code{Quot.sound}. Unexpected assumptions prevent issuance. The proof must
also emit an audited completion marker specific to the input digest. A
successful process exit alone is insufficient. Source guards separately check
for proof placeholders in the project library.

\begin{table}[t]
\centering
\begin{tabular}{p{0.24\textwidth}p{0.68\textwidth}}
\hline
Artifact & Purpose \\
\hline
\code{inputs.json} & Canonical mathematical inputs and proposed result, where applicable. \\
\code{Proof.lean} & Generated statements and proofs for the request. \\
\code{Proof.olean} & Compiled checked proof artifact. \\
\code{transcript.log} & Backend commands, environment checks and proof output. \\
\code{certificate.json} & Claim scope, assumptions, versions and artifact hashes. \\
\hline
\end{tabular}
\caption{Contents of a formal certificate bundle. A program's parent run report
records provenance but is not an additional mathematical certificate.}
\label{tab:certificate}
\end{table}

\code{statclaim check-certificate bundle} validates the bundle, reconstructs
derived fields and regenerates the expected proof statements for its fixed
scope. Replay requires the bound source and dependency versions and reruns the
proof and audit. Changing metadata cannot turn a design certificate into a
proof of an observed p-value or a stronger error bound. The old transcript and
artifact hashes are provenance records. Replay obtains mathematical evidence
by checking the retained proof against the bound definitions again.

The trust boundary includes the \proglang{Lean} kernel and imported artifacts,
the statistical meaning of the definitions, and the application code translating
declarations into statements and binding them to reports. Hashes detect content
changes but do not authenticate authorship and bundles are not digitally signed.
The independent derivation checker, regression tests and replay checks examine
parts of this boundary without constituting a formal verification of the
entire software stack.

\section{Planning and operating characteristics} \label{sec:planning}

\subsection{Targets and candidate designs}

Planning specifies an analysis, parameter values and a target. A sample-size
search also specifies a set $\mathcal{D}$ of candidate designs. Each candidate
$d\in\mathcal{D}$ has a sampling plan and an associated test. Let
$P_{\theta,d}$ be the probability law of the recorded data under candidate $d$
at parameter value $\theta$, and let $R_d$ be the event that its test rejects
the null. A planning point $\theta_1$ specifies an alternative effect and any
other parameters needed to calculate rejection probabilities, such as a
baseline response rate or an outcome variance. Power at this point is
\begin{equation}
 \pi(d,\theta_1)=P_{\theta_1,d}(R_d).
 \label{eq:power}
\end{equation}
A power target specifies a minimum power $1-\beta$ and a maximum Type~I error
$\alpha_0$, where $0<\beta<1$ and $0\leq\alpha_0\leq1$. Here $\beta$ is the
maximum probability of failing to reject at the planning point. A candidate
must satisfy $\pi(d,\theta_1)\geq1-\beta$ and the requested error bound under
its declared null. The null hypothesis and planning alternative are specified
separately. StatClaim rejects a point inside the null when an alternative is
required. The power claim applies at the specified planning point.

After importing \code{planning}, the introductory Fisher analysis can be
extended with a sample-size search.
\begin{CodeInput}[samepage=true]
let alternative = binomial_rates(p_control=1/10, p_treatment=2/5)
let target = power_target(at=alternative, minimum=4/5,
                          maximum_typeI=1/20)
let candidates = equal_arm_sizes(analysis=analysis,
                                 min_per_arm=2, max_per_arm=40)
let sizing = smallest_sample_size(candidates=candidates,
                                  target=target)
let recommendation = calculate(sizing)
let sizing_verification = verify(sizing)
\end{CodeInput}
\studiolink{paper/fisher_sample_size}{Open the complete program in StatClaim Studio.}
The planning point sets the control and treatment response rates to 0.1 and
0.4. The target requires at least 80\% power and a Type~I error bound of at
most 5\%. The smallest sufficient size in the declared range is 36 per arm.
The calculation selects this size. Verification checks that it meets the
target and that every smaller candidate in the range fails the target.
Minimality applies to equal allocation with 2 to 40 participants per arm.
Each smaller candidate is checked separately because the power of a discrete
test need not increase with sample size.

\code{sample\_size} requests a sufficient candidate.
\code{smallest\_sample\_size} also requests minimality where an adapter supports
it. Sizes can refer to participants, clusters, matched pairs or events,
depending on the design. If no candidate meets the target, the search reports
that no sufficient size was found in the declared range. The analysis level
and the target's error limit $\alpha_0$ are checked separately. When the error
theorem bounds Type~I error by the analysis level, using that bound to meet
the target requires a level at most $\alpha_0$.

\subsection{Exact, approximate and adaptive planning}

Exact binomial planning sums rejection probabilities at rational planning
points. Supported formal adapters check the corresponding power and error
claims in \proglang{Lean}. Two-stage single-arm designs specify interim and
final rules and evaluate the probability of rejection under the entire policy.
Count rules compare response counts with declared cutoffs. Posterior rules
specify a beta prior with positive integer parameters and a reference response
rate. They continue or reject when the posterior probability of exceeding
that rate reaches a declared threshold. Type~I error and power are evaluated
over the possible data and stopping decisions under the sampling model. The
formal posterior rule uses a finite binomial sum to define this probability.
The library does not prove its equality to the corresponding beta-distribution
integral.

Searches for Simon-type two-stage designs \citep{Simon:1989} can minimize
expected sample size under the null or the largest possible sample size among
the declared candidates. The current certificate checks that the selected
design meets its error and power targets. The ranking of candidates by sample
size is calculated and has no formal optimality certificate.

Cited methods use planning formulas such as normal approximations or
calculations with a noncentral t distribution. Their power estimates are
reported as \code{calculated}, with the assumptions of the planning method.
StatClaim does not formally verify the accuracy of these approximations.
Unsupported planning operations produce a diagnostic.

\subsection{Simulation and its uncertainty}

Simulation generates independent studies under a specified data distribution
and applies the declared test to each study. The proportion of studies that
reject estimates the rejection probability under that scenario. Monte Carlo
uncertainty describes the uncertainty due to the finite number of simulated
studies \citep{Morris+White+Crowther:2019}. The following self-contained program
compares a power formula with a simulation estimate for a pooled two-sample
t-test.
\begin{CodeInput}[samepage=true]
use models.two_arm_continuous
use sampling
use tests.t
use planning

let study = two_arm_continuous(n_control=40, n_treatment=40)
let analysis = t_test(design=fixed(model=study),
                      null=equal_means(model=study), alpha=1/20)
let target = power_target(at=mean_difference(value=5, sd=10),
                          minimum=4/5, maximum_typeI=1/20)
let formula = calculate(meets(analysis=analysis, target=target))
let estimate = simulate(meets(analysis=analysis, target=target),
                         seed=2026, runs=10000)
\end{CodeInput}
\studiolink{paper/t_test_simulation}{Open this example in StatClaim Studio.}
This scenario has normally distributed outcomes with a mean difference of 5
and a common standard deviation of 10. Each arm has 40 participants, and the
simulation uses 10,000 studies. Rejection proportions are reported with 95\%
Clopper--Pearson intervals \citep{Clopper+Pearson:1934}. The power component of
the target passes when the interval's lower endpoint is at least the requested
minimum. It fails when the upper endpoint is below that minimum. Otherwise,
the power assessment remains undecided.

A supported sample-size search examines candidates and confirms a passing
candidate using an independent random stream. The result reports a size that
passes this assessment. It provides no proof that this size is the smallest
one with sufficient true power. Seeds, run counts, scenario details and the
simulation engine version are retained for reproduction.

Supported scenarios include unequal variances, skewed outcomes, unequal
cluster sizes and delayed treatment effects in survival studies. Designs with
planned interim analyses are simulated stage by stage. Stopping probabilities
and expected sample size are reported for the complete procedure.

Blinded sample-size re-estimation updates the planned size using an interim
variance estimate from the pooled treatment groups. The current adapter
assesses its error rate at explicitly supplied null scenarios. When Type~I
error is assessed by simulation, the upper endpoint of its interval must be
at most $\alpha_0$ at every supplied null point for that component of the target
to pass. The assessment applies to those points.

Family simulations can estimate the probability of rejecting at least one
endpoint or of rejecting all endpoints. They also estimate familywise error
under the global null, where every endpoint's null hypothesis is true.
Strong familywise error control requires a bound over every configuration of
true and false null hypotheses.

\subsection{Precision targets}

Precision planning specifies an interval procedure, a confidence level, a
planning point and a width criterion. For an interval $[L,U]$, where $L$ and
$U$ are its lower and upper endpoints, the absolute margin of error is
$(U-L)/2$. For proportions and differences, the relative margin is
$(U-L)/(2|q|)$, where $q\ne0$ is the planning value of the quantity being
estimated. For ratios, the relative margin is $1-\sqrt{L/U}$, computed from
the interval's width on the logarithmic scale. This criterion requires finite
endpoints with $0<L\leq U$. A ratio interval with a nonpositive lower endpoint
or an infinite upper endpoint cannot meet a finite relative-precision target.

An optional assurance sets a minimum probability that the eventual interval
meets the width criterion under the planning point. The confidence level
concerns coverage of the unknown parameter, as defined in
Equation~\ref{eq:coverage}.

The following program specifies precision for a one-arm binomial response
rate.
\begin{CodeInput}[samepage=true]
use tests.binomial
use planning

let arm = one_arm_binomial(n=60)
let interval = planned_interval(model=arm, measure="response_rate")
let target = precision_target(at=response_rate(value=3/10),
                              level=19/20, margin_of_error=1/10)
let sizes = interval_sizes(interval=interval,
                           min_size=10, max_size=2000)
let needed = calculate(smallest_sample_size(candidates=sizes,
                                           target=target))
\end{CodeInput}
\studiolink{paper/response_rate_precision}{Open this example in StatClaim Studio.}
The target uses a planning response rate of 0.3, a 95\% confidence level and
an absolute margin of error of at most 0.1. The search considers 10 to 2,000
participants. Without assurance, precision is evaluated at the counts expected
under the planning point. With assurance, supported count distributions are
enumerated to calculate the probability of meeting the width criterion. For
designs that fix only a total sample size, this calculation uses the expected
group allocation where documented.

The result identifies whether the interval method is exact or uses a normal
approximation. All current precision searches are numerical calculations,
including those based on exact interval methods. A formal coverage guarantee
for the interval procedure does not verify the numerical precision search.

\section{Authoring interfaces and extension} \label{sec:interfaces}

\subsection{Command line and editor integration}

The command line supports compilation, explanations, reports and
execution. \code{compile} checks declarations, and \code{explain} shows the
contract derivations. \code{report} summarizes claims, assumptions and
requirements. \code{check} executes the requested calculations, simulations
and verification. Structured JSON output supports programmatic inspection.
The following commands illustrate the workflow for a file named
\code{study.stat}.
\begin{Code}[samepage=true]
statclaim compile study.stat
statclaim explain study.stat
statclaim report study.stat
statclaim check study.stat --output run
statclaim check-certificate run/request-001
\end{Code}
The output directory \code{run} must not exist before execution. If the first
request issues a formal certificate, the last command rechecks its bundle in
\code{run/request-001}. Each result states its scope and lists the
files produced by the request.

The language server uses the Language Server Protocol (LSP) to provide
diagnostics, completion, hover information and source navigation in
\pkg{Visual Studio Code} \citep{VSCode}. Completion uses the imported
constructors, their accepted arguments and the public fields of language
objects. A diagnostic identifies the problematic argument and, where
available, the declaration that supplies its value. For an incompatible null,
it can identify both the hypothesis declaration and the analysis using it.

The extension's Claims view lists declared claims, open requirements and
execution results. Results display their assumptions, evidence grades and
available certificates. Each execution result is linked to the document
version and a hash of the source text used for that run. Editing the file or
restarting the server marks earlier results as outdated. These results can
remain visible as history, with their original source version.

\subsection{StatClaim Studio} \label{sec:studio}

StatClaim Studio is a graphical block editor for the same
\code{.stat} language. A guided new-study workflow asks the researcher to
choose a design and supply its inputs. It checks that the generated
specification compiles before creating the study. On the canvas, blocks
represent constructor calls and references to named values. Connectors show
the types accepted by each input. Parameters can be edited directly, and the
textual source remains available for inspection. Saving produces a
\code{.stat} file that can also be run from the command line.

The \href{https://ssm123ssm.github.io/statClaim/}{web Studio} opens the complete
programs linked throughout this article. It runs compilation, calculations,
simulations and cited-method checks in the browser. Formal proof generation
requires \proglang{Lean} and is unavailable in the browser. To verify the same
programs, use the full Studio in the
\href{https://codespaces.new/ssm123ssm/statClaim?quickstart=1}{project's GitHub Codespace}
or a local installation with the formal dependencies described in the
computational details. Table~\ref{tab:paper-studio} indexes the article's
programs and their Studio links.

The server uses the shared registry of language objects and constructor
definitions to obtain argument types, available choices and public fields.
It generates compatibility hints by trying the selected value as an input,
with representative values for the remaining arguments. For example, a
continuous-outcome study can suggest a t-test. Fisher's binomial analysis is
incompatible with that outcome type. Unsupported combinations receive the
compiler's diagnostic. Suggested replacement tests use the same declared null
hypothesis.

The parser records source ranges for declarations and arguments. Studio uses
these ranges to translate block edits into source edits. The common compiler
checks each edit. Run controls use the same execution layer as the command
line, and results show assumptions, calculations, evidence grades and
available certificates. Changing a value marks earlier results as outdated.
Direct parameter exploration can automatically rerun calculations and
simulations. Formal proofs remain attached to the previous source until
verification is requested again.

The distribution includes a guided tutorial with 45 lessons and corresponding
\code{.stat} files. The lessons cover declarations and diagnostics, exact and
cited methods, composition, planning and simulation. The tutorial can be read
in a browser, and its source files can be used from the command line, the
editor or Studio. Some examples deliberately produce diagnostics or leave
requirements unresolved. Appendix~\ref{app:tutorials} summarizes the topics,
representative results and software requirements. Additional editable designs
are supplied in \path{examples/}.

\subsection{Adding methods and maintaining consistency}

Methods are added through registration in the repository. External plugin
loading is not currently supported. A method entry supplies typed
constructors, supported outcome and null forms, assumptions and guarantees.
Functions for calculation and simulation are registered where available. The
registry also supplies documentation and descriptions used by the interfaces.
The classified catalogue groups public objects and constructors by role.

Formal verification support requires mathematical definitions and proofs.
An adapter translates supported language declarations into statements using
those definitions. Certificate issuance and replay must check the statements
for the requested claim. Tests for a new method should cover compatible and
incompatible declarations, unmet requirements, source locations and concrete
calculations. Methods with proof support also need tests of proof checking
and rejection of altered certificates. Review must check that the formal
definitions and assumptions describe the intended study and that the theorem
establishes the requested claim.

\section{Validation and application} \label{sec:validation}

The replication materials include a check of published two-stage designs and
an encoding of selected components of a clinical trial's statistical analysis
plan. The first examines numerical agreement and formal guarantees. The second
examines how a protocol with several hypotheses is represented and which
evidence is available for its claims.

\subsection{Rechecking Simon's two-stage designs}

\citet{Simon:1989} develops single-arm phase~II trial designs with a binary
response endpoint. These designs allow early stopping when a treatment shows
insufficient activity while meeting specified Type~I error and power targets.
We use the published designs to assess StatClaim's calculations and formal
verification of complete two-stage policies.

Tables 1 and 2 of the paper contain \simonDesigns{} entries covering
optimal and minimax designs for 51 parameter settings. Each design specifies
a first-stage size $n_1$, a maximum total size $n$, an interim cutoff $r_1$
and a final cutoff $r$. The study stops for futility when the first-stage
response count is at most $r_1$. Otherwise, it continues to $n$ participants
and rejects the null when the total response count exceeds $r$. The null
boundary $p_0$ and planning alternative $p_1$ are response probabilities, with
$p_1>p_0$. The requested Type~I error limit is $\alpha$ and the minimum power
at $p_1$ is $1-\beta$. Optimal designs minimize expected sample size at $p_0$.
Minimax designs minimize the maximum total sample size.

For every entry, we encode the stopping rule, recalculate its operating
characteristics and obtain \proglang{Lean} proofs of its error and power
bounds. We also check that numerical design searches reproduce the tabulated
designs and compare the calculations with an independent \proglang{R}
implementation. The proofs cover the operating bounds of each selected
design. Optimality is assessed by numerical search.

The first tabulated optimal design has $n_1=9$, $n=24$, $r_1=0$ and $r=2$,
with $p_0=0.05$, $p_1=0.25$ and $\alpha=\beta=0.10$. Its StatClaim
specification is
\begin{CodeInput}[samepage=true]
use designs.two_stage
use planning

let stages = two_stage_binomial(n_stage1=9, n_stage2=15)
let rule = threshold_rule(continue_above=0, reject_above=2)
let analysis = two_stage_test(design=stages, rule=rule,
                              null_rate=1/20)
let target = power_target(at=response_rate(value=1/4),
                          minimum=9/10, maximum_typeI=1/10)
let checked = verify(meets(analysis=analysis, target=target))
\end{CodeInput}
\studiolink{paper/simon_design}{Open this example in StatClaim Studio.}
The second-stage size is 15 because the total size is 24. The exact
calculation gives rejection probability \simonExampleError{} at $p_0$ and
power \simonExamplePower{} at $p_1$. Expected sample size at $p_0$ is
\simonExampleExpected{}. Verification establishes the error bound for every
response probability $p\in[0,p_0]$ and the power bound at $p_1$.

StatClaim can also select the sample sizes and stopping cutoffs from the
planning target. Continuing the program above, the following searches compare
designs with a maximum total size of 120 participants.
\begin{CodeInput}[samepage=true]
let optimal_candidates = two_stage_designs(null_rate=1/20,
                            max_total=120, criterion="optimal")
let optimal = calculate(sample_size(candidates=optimal_candidates,
                                    target=target))
let minimax_candidates = two_stage_designs(null_rate=1/20,
                            max_total=120, criterion="minimax")
let minimax = calculate(sample_size(candidates=minimax_candidates,
                                    target=target))
\end{CodeInput}
\studiolink{paper/simon_search}{Open the complete search program in StatClaim Studio.}
The selected designs are
\begin{center}
\begin{tabular}{lrrrrr}
\hline
Criterion & $n_1$ & $n$ & $r_1$ & $r$ & Expected size at $p_0$ \\
\hline
Optimal & 9 & 24 & 0 & 2 & 14.546 \\
Minimax & 13 & 20 & 0 & 2 & 16.407 \\
\hline
\end{tabular}
\end{center}
The optimal design has the smaller expected sample size under the null. The
minimax design has the smaller maximum total size. Both searches return
\code{calculated} results. Replacing \code{calculate} with \code{verify}
checks the selected design's error and power bounds in \proglang{Lean}.
The ranking by the chosen criterion remains numerical.

Table~\ref{tab:simon-validation} summarizes the checks across all entries.
Exact calculations meet every stated error and power target. The numerical
search over count-threshold designs with at most 120 participants returns
every tabulated design. The formal certificates establish the operating
targets of the selected designs. The ranking of candidates in the search is
computed numerically and is not covered by an optimality proof.

\begin{table}[!htb]
\centering
\small
\begin{tabular}{p{0.58\textwidth}rp{0.23\textwidth}}
\hline
Check & Entries & Basis \\ \hline
Error and power targets met & 102/102 & Exact calculation \\
Error and power bounds checked in Lean & 102/102 & Formal proof \\
Bounded search returns the tabulated design & 102/102 & Calculation \\
\pkg{clinfun} returns the tabulated design & 102/102 & Comparison \\
Printed early-stopping probability agrees & 98/102 & Comparison \\
\hline
\end{tabular}

\caption{Checks of the \simonDesigns{} entries in Simon's tables. Each entry
has both an error and a power claim. Formal proofs concern these claims for
the complete stopping policy. Search and comparison results are numerical.}
\label{tab:simon-validation}
\end{table}

The comparison script also runs \code{ph2simon} from the \proglang{R} package
\pkg{clinfun} \citep{clinfun} with the same maximum size of 120. It returns the
same designs in all cases. Expected sample sizes agree within 0.005 and
early-stopping probabilities within 0.00005. All printed expected sample
sizes agree with the recalculations at the precision used in the replication.
Four printed early-stopping probabilities differ from the recalculations by
more than 0.005. The recalculated probabilities agree with \pkg{clinfun}.

\begin{samepage}
\subsection{A COPD trial analysis plan}

The second example encodes selected components of the AERISTO statistical
analysis plan, edition 2.0, dated 8 June 2018, for study NCT03162055
\citep{AERISTO:2018}. This is a trial in chronic obstructive pulmonary disease
(COPD). The plan specifies 500 participants per arm and anticipates about
440 per arm in its per-protocol analyses. The encoding covers the principal
testing hierarchy and four planning statements.
\end{samepage}

The primary family has two non-inferiority hypotheses and one superiority
hypothesis. Its one-sided error budget is 2.5\%, initially split equally
between the two non-inferiority tests. Rejections allow the allocated level
to pass around a declared graph. If all three hypotheses are rejected, testing
proceeds through four secondary endpoints in a fixed order at 2.5\%.

The file \path{replication/copd/design_cited.stat}, available in
\studiolink{copd/design_cited}{StatClaim Studio}, uses cited mixed-model
contracts for continuous outcomes and a cited logistic-regression contract for
the binary outcome. A \code{graphical} declaration represents the primary
family. A \code{fixed\_sequence} declaration places this family before the
four secondary analyses. Table~\ref{tab:copd-evidence} records the resulting
evidence. The primary family's guarantee is cited and has large-sample scope.
The sequence has a combination theorem checked by \proglang{Lean} conditional on its
members' guarantees, with the same inherited large-sample scope. The weakest
evidence in the file is cited. A
\studiolink{copd/design_declared}{second encoding} uses author-declared tests
and reports declared evidence.

\begin{table}[!htb]
\centering
\small
\begin{tabular}{p{0.24\textwidth}p{0.45\textwidth}p{0.21\textwidth}}
\hline
Component & Representation & Result \\
\hline
Individual tests & Mixed-model and logistic-regression contracts & Cited \\
Primary family & Graphical allocation and recycling & Cited \\
Complete hierarchy & Fixed sequence beginning with the primary family & Conditional \\
Planning statements & Numerical power calculations & Calculated \\
\hline
\end{tabular}
\caption{Evidence for the encoded components of the COPD analysis plan.
The complete hierarchy's result assumes its members' guarantees.}
\label{tab:copd-evidence}
\end{table}

The file \path{replication/copd/power.stat}, also available in
\studiolink{copd/power}{StatClaim Studio}, evaluates the four statements in
Section 1.3 of the plan using the differences and standard deviations it
supplies. Forced expiratory volume in one second is denoted by FEV$_1$, and
TDI denotes the Transition Dyspnea Index. For FEV$_1$, the plan gives
effective standard deviations for the change from baseline over 24 weeks.
Table~\ref{tab:copd-power} shows the plan's power statements and the calculated
values. The last row uses the plan's additional planning scenario with a
margin of $-0.75$ TDI units. Its one-sided level is taken as 2.5\%, the level
specified for the secondary hierarchy.

\begin{table}[!htb]
\centering
\small
\begin{tabular}{p{0.51\textwidth}p{0.22\textwidth}r}
\hline
Statement & Plan's statement & Calculated power \\ \hline
Trough FEV$_1$ non-inferiority, 2.5\% & about 94\% & 94.44\% \\
Trough FEV$_1$ non-inferiority, 1.25\% & 90\% & 90.51\% \\
Peak FEV$_1$ superiority, 1.25\% & at least 97\% & 97.52\% \\
TDI non-inferiority, margin $-0.75$ & at least 90\% & 95.98\% \\
\hline
\end{tabular}

\caption{Individual-test power calculations for the COPD plan. These are
normal approximations using the supplied planning inputs. They assess each
listed test at its stated level.}
\label{tab:copd-power}
\end{table}

The calculated powers agree with all four statements. For trough FEV$_1$ at
1.25\%, the calculated rejection threshold is
$\copdCriticalTroughB$~mL. The plan mentions $-24.5$~mL, a difference of about
0.26~mL. These calculations concern individual tests. Joint power of the
complete testing hierarchy is not assessed here. The standard deviations are
planning inputs, and the reports retain the model and
missing-data assumptions used by the cited methods.

The scope of this encoding is limited to the declared components. Detailed
covariates, covariance specifications, fallback analyses and analysis-set
membership rules are not encoded. Additional superiority tests that do not determine progression through
the secondary hierarchy are also omitted. The margin of $-0.1$ for the Early
Morning Symptoms COPD Instrument (EMSCI) severity score is encoded as written,
which treats higher scores as better. Neither the plan nor the protocol states
this direction, and the reports list it as an assumption. If higher scores
indicate more severe symptoms, the non-inferiority null would instead be a
treatment difference of at least $0.1$. The familywise claim concerns the
encoded sequence. Agreement between the protocol and analysis plan was not
checked, and no observed trial outcomes were analysed.

\section{Summary and discussion} \label{sec:summary}

StatClaim is a declarative domain-specific language for expressing study
specifications as programs. Models, sampling plans, hypotheses and analyses are
declared as typed objects, and contracts check their compatibility. For
supported verification requests, StatClaim generates proof obligations and
invokes \proglang{Lean} to construct and check proofs.

Researchers can use StatClaim during protocol development to calculate sample
size, power and precision for proposed designs. These planning requests refer
to declared models and analyses, so their inputs and assumptions remain part
of the study specification. When a design is revised, recompilation
checks the updated declarations, and subsequent execution evaluates the revised
claims. The program and its results can therefore accompany a protocol or
statistical analysis plan as a record of what was specified and checked.
Reviewers can inspect this record through the command line, editor or Studio,
and replay formal certificates against their recorded inputs and dependencies.

A study specification may need methods for which formal proofs are not yet
available. StatClaim accommodates these methods through evidence grades that
identify the basis of each guarantee. For example, a conditional composition
proof establishes a family guarantee assuming the guarantees of its component
tests. Those component guarantees may themselves be formally proved, cited
from the literature or declared by the author.

The examples in Section~\ref{sec:validation} illustrate both these uses. The Simon
replication selects two-stage sample sizes and stopping rules and verifies
their error and power bounds. The COPD example encodes part of a published
analysis plan using cited methods and a conditional composition proof.

The interpretation of these results depends on the correspondence between the
encoded study and the study conducted. StatClaim does not establish that a
declared sampling model describes the recruited population or that an endpoint
was selected before outcomes were examined. Researchers must justify these
choices and assess whether the recorded assumptions hold. Formal verification
also depends on the definitions representing the intended statistical
procedure. The translation from declarations to proof statements and the
verification toolchain remain part of the trust boundary described in
Section~\ref{sec:verification}.

Within this scope, the current implementation has limits in coverage and
computation. Formal verification is available for selected procedures, and
exact enumeration and kernel computation restrict the sizes some adapters can
handle. StatClaim also supports approximate planning, simulation and precision
searches, whose numerical results have the scope described in
Section~\ref{sec:planning}. 

Extending this coverage will require new proofs for interval procedures,
precision searches and adaptive sampling policies. Each extension needs an
explicit observation law and a precise statement of the guarantee to be
established. Continuous and
large-sample methods will require further distributional and asymptotic
arguments. Statistical and formal review of new modules will be needed to
check that these definitions and proofs support their advertised claims.

As the method library grows, reusable procedure components and a versioned
extension interface could make it easier to add and maintain methods. Future
work will explore refinement types and dependent types to express shared requirements,
such as model identity and sampling policies, more precisely in the DSL's
contract layer. The formal backend already uses
\proglang{Lean}'s dependent type system to represent statements and check
proofs. Planned development also includes an \proglang{R} package for invoking StatClaim
specifications directly from \proglang{R} scripts, allowing researchers to
incorporate design checks and planning calculations into their existing
workflows.

StatClaim is an open-source project released under the MIT license and
intended to grow through community contributions. Its modular method
definitions, shared registry and separate formal adapters are structured to
support extensions to the language, statistical methods and authoring tools.
The contribution guide documents how to add and test these extensions,
providing a common starting point for statisticians, formal-methods
researchers and software developers.

\section*{Computational details}

The executable examples were checked with \proglang{Python} 3.12.15,
\pkg{statclaim} 0.1.0 (language version 0.41.0) and \proglang{Lean} 4.35.0.
Python dependency pins are recorded in \code{uv.lock}. Formal toolchain and
library pins appear in \code{lean-toolchain} and \code{lake-manifest.json}. Formal
verification and certificate replay require the project checkout and its
formal dependencies. From that checkout, \code{make setup} prepares the
supported development environment. The commands in the text can then be run
with the prefix \code{uv run -{}-locked}. Runtime reports retain source digests
and backend versions because the package version alone does not identify a
particular development snapshot. The companion script runs the displayed
examples and checks the diagnostic variants. With \code{-{}-lean}, it also
verifies the formal requests, compares the printed outputs and replays an
observed-result certificate:
\begin{Code}[samepage=true]
uv run --locked python \
  statClaim_paper/manuscript/check_examples.py --lean
\end{Code}

The validation results in Section~\ref{sec:validation} are reproduced by
\code{make replicate} for calculations and stored-output checks, and by
\code{make replicate-lean} for fresh formal proofs. The Simon comparison with
\pkg{clinfun} uses the separate \proglang{R} script
\path{replication/comparison/simon_clinfun.R}. Its package requirements and
recorded output are supplied in the same directory. The manuscript's
validation tables and numerical values are generated with
\begin{Code}[samepage=true]
uv run --locked python \
  statClaim_paper/manuscript/make_tables.py --validation
\end{Code}

\bibliography{statclaim}

@Misc{AERISTO:2018,
  author       = {{AstraZeneca}},
  title        = {Statistical Analysis Plan, Study {D5970C00002} ({AERISTO}): A Randomised,
                  Double-Blind, Double-Dummy, Multicentre, Parallel Group Study to Assess the
                  Efficacy and Safety of Glycopyrronium/Formoterol Fumarate Fixed-Dose
                  Combination Relative to Umeclidinium/Vilanterol Fixed-Dose Combination over 24
                  Weeks in Patients with Moderate to Very Severe Chronic Obstructive Pulmonary
                  Disease},
  year         = {2018},
  note         = {Edition 2.0, 8 June 2018. Posted with ClinicalTrials.gov record
                  NCT03162055},
  url          = {https://cdn.clinicaltrials.gov/large-docs/55/NCT03162055/SAP_001.pdf},
}

@Manual{statclaim,
  title        = {\pkg{statclaim}: Typed Statistical Claims with Graded Evidence},
  author       = {Supun Manathunga},
  year         = {2026},
  note         = {\proglang{Python} package version 0.1.0},
  url          = {https://github.com/ssm123ssm/statClaim},
}

@Article{Blair+Cooper+Coppock:2019,
  author       = {Graeme Blair and Jasper Cooper and Alexander Coppock and Macartan Humphreys},
  title        = {Declaring and Diagnosing Research Designs},
  journal      = {American Political Science Review},
  year         = {2019},
  volume       = {113},
  number       = {3},
  pages        = {838--859},
  doi          = {10.1017/S0003055419000194},
}

@InProceedings{Kawamoto+Kobayashi+Suenaga:2025,
  author       = {Yusuke Kawamoto and Kentaro Kobayashi and Kohei Suenaga},
  title        = {\pkg{StatWhy}: Formal Verification Tool for Statistical Hypothesis Testing
                  Programs},
  booktitle    = {Computer Aided Verification -- {CAV} 2025},
  year         = {2025},
  series       = {Lecture Notes in Computer Science},
  volume       = {15932},
  pages        = {216--230},
  publisher    = {Springer},
  doi          = {10.1007/978-3-031-98679-6_10},
}

@Manual{clinfun,
  title        = {\pkg{clinfun}: Clinical Trial Design and Data Analysis Functions},
  author       = {Venkatraman Seshan and Karissa Whiting},
  year         = {2026},
  note         = {\proglang{R} package version 1.1.6},
  doi          = {10.32614/CRAN.package.clinfun},
}

@Manual{DeclareDesign,
  title        = {\pkg{DeclareDesign}: A Tidyverse-First Research Design Declaration and
                  Diagnosis Tool},
  author       = {Alexander Coppock and Graeme Blair and Jasper Cooper and Macartan Humphreys
                  and Neal Fultz},
  year         = {2026},
  note         = {\proglang{R} package version 2.0.0},
  url          = {https://declaredesign.org/r/declaredesign/},
}

@Manual{R,
  title        = {\proglang{R}: A Language and Environment for Statistical Computing},
  author       = {{R Core Team}},
  organization = {R Foundation for Statistical Computing},
  address      = {Vienna, Austria},
  year         = {2026},
  doi          = {10.32614/R.manuals},
}

@Article{Simon:1989,
  author       = {Richard Simon},
  title        = {Optimal Two-Stage Designs for Phase {II} Clinical Trials},
  journal      = {Controlled Clinical Trials},
  year         = {1989},
  volume       = {10},
  number       = {1},
  pages        = {1--10},
  doi          = {10.1016/0197-2456(89)90015-9},
}

@InProceedings{deMoura+Ullrich:2021,
  author       = {Leonardo de Moura and Sebastian Ullrich},
  title        = {The \proglang{Lean} 4 Theorem Prover and Programming Language},
  booktitle    = {Automated Deduction -- {CADE} 28},
  year         = {2021},
  pages        = {625--635},
  publisher    = {Springer-Verlag},
  doi          = {10.1007/978-3-030-79876-5_37},
}

@Manual{Python,
  author       = {{Python Software Foundation}},
  title        = {\proglang{Python} 3 Documentation},
  year         = {2026},
  url          = {https://docs.python.org/3/},
}

@InProceedings{mathlib2020,
  author       = {{The mathlib Community}},
  title        = {The \proglang{Lean} Mathematical Library},
  booktitle    = {Proceedings of the 9th ACM SIGPLAN International Conference
                  on Certified Programs and Proofs},
  year         = {2020},
  publisher    = {ACM},
  doi          = {10.1145/3372885.3373824},
}

@Article{Morris+White+Crowther:2019,
  author       = {Tim P. Morris and Ian R. White and Michael J. Crowther},
  title        = {Using Simulation Studies to Evaluate Statistical Methods},
  journal      = {Statistics in Medicine},
  year         = {2019},
  volume       = {38},
  number       = {11},
  pages        = {2074--2102},
  doi          = {10.1002/sim.8086},
}

@Article{Clopper+Pearson:1934,
  author       = {C. J. Clopper and E. S. Pearson},
  title        = {The Use of Confidence or Fiducial Limits Illustrated in the
                  Case of the Binomial},
  journal      = {Biometrika},
  year         = {1934},
  volume       = {26},
  number       = {4},
  pages        = {404--413},
  doi          = {10.1093/biomet/26.4.404},
}

@Manual{VSCode,
  author       = {{Microsoft}},
  title        = {Language Server Extension Guide},
  note         = {\pkg{Visual Studio Code} documentation},
  year         = {2026},
  url          = {https://code.visualstudio.com/api/language-extensions/language-server-extension-guide},
}

@InProceedings{Freeman+Pfenning:1991,
  author       = {Tim Freeman and Frank Pfenning},
  title        = {Refinement Types for \proglang{ML}},
  booktitle    = {Proceedings of the ACM SIGPLAN 1991 Conference on Programming
                  Language Design and Implementation},
  year         = {1991},
  pages        = {268--277},
  publisher    = {ACM},
  doi          = {10.1145/113445.113468},
}

@InProceedings{Rondon+Kawaguchi+Jhala:2008,
  author       = {Patrick M. Rondon and Ming Kawaguchi and Ranjit Jhala},
  title        = {Liquid Types},
  booktitle    = {Proceedings of the 2008 ACM SIGPLAN Conference on Programming
                  Language Design and Implementation},
  year         = {2008},
  pages        = {159--169},
  publisher    = {ACM},
  doi          = {10.1145/1375581.1375602},
}

\clearpage
\appendix
\section{Tutorial and example coverage} \label{app:tutorials}

The software archive includes a guided tutorial in \path{docs/tutorial.html}
and runnable sources in \path{examples/tutorial/}. There are 45 numbered
lessons and 46 source files because Lesson 3 has separate examples for an
incompatible null and unspecified adaptive stopping.
Table~\ref{tab:tutorial-coverage} groups the lessons by topic.

\begin{table}[!htb]
\centering
\begin{tabular}{p{0.12\textwidth}p{0.79\textwidth}}
\hline
Lessons & Topics \\
\hline
1--8 & Declarations, claims, result fields, diagnostics, observed data,
multiplicity, sample-size searches and non-inferiority. \\
9--18 & Single-arm and two-stage designs, posterior stopping rules, cited
methods, interim analyses, composition and approximate power. \\
19--27 & Shared assignments, clusters, repeated measurements, randomization,
analysis populations and observational designs. \\
28--36 & Simulated power, observational and trial sample sizes, regression,
survey planning and interval precision. \\
37--45 & Simulation with unequal clusters, skewness and delayed effects,
interim analyses, size re-estimation, family power and exact planning. \\
\hline
\end{tabular}
\caption{Topic coverage of the guided tutorial.}
\label{tab:tutorial-coverage}
\end{table}

Table~\ref{tab:tutorial-results} identifies selected requests and the results
to inspect. The results refer to the supplied files. Some lessons deliberately
produce a diagnostic or an unresolved requirement. The execution environments
are described in Section~\ref{sec:studio}.

\begin{table}[!htb]
\centering
\small
\begin{tabular}{p{0.43\textwidth}p{0.35\textwidth}p{0.12\textwidth}}
\hline
Source file & Selected result & Lean \\
\hline
\code{01\_first\_claim.stat} & Verified Type~I error bound & Yes \\
\code{03\_wrong\_null.stat} & Incompatible hypothesis diagnostic & No \\
\code{03\_optional\_stopping.stat} & Unresolved sampling requirement & No \\
\code{07\_sample\_size.stat} & Verified minimal sample size & Yes \\
\code{12\_continuous\_outcome.stat} & Cited test guarantee & No \\
\code{13\_mixed\_endpoints.stat} & Conditional family guarantee & Yes \\
\code{28\_simulated\_power.stat} & Calculated and simulated power & No \\
\code{36\_precision.stat} & Calculated interval precision & No \\
\hline
\end{tabular}
\caption{Representative tutorial requests and their computational
requirements. File names are relative to \code{examples/tutorial/}. The
conditional family in Lesson 13 has cited evidence among its members.}
\label{tab:tutorial-results}
\end{table}

The files use the compilation and execution commands described in
Section~\ref{sec:interfaces}. They can also be opened in the editor or imported
into Studio. The generated reference in \path{docs/catalogue.md} records
constructor availability, arguments and restrictions for the distributed
software version.

\clearpage
Table~\ref{tab:paper-studio} lists the complete programs used in the main text.
Each link opens the corresponding source in the web Studio. The source files
are also included in the replication materials.

\begin{table}[!htb]
\centering
\small
\begin{tabular}{p{0.27\textwidth}p{0.55\textwidth}l}
\hline
Example & Source file & Studio \\
\hline
Fisher design guarantee & \code{examples/paper/fisher\_design.stat} & \studiolink{paper/fisher_design}{Open} \\
Unspecified stopping & \code{examples/paper/fisher\_adaptive.stat} & \studiolink{paper/fisher_adaptive}{Open} \\
Model composition & \code{examples/paper/composed\_model.stat} & \studiolink{paper/composed_model}{Open} \\
Observed Fisher result & \code{examples/paper/fisher\_observed.stat} & \studiolink{paper/fisher_observed}{Open} \\
Fisher sample size & \code{examples/paper/fisher\_sample\_size.stat} & \studiolink{paper/fisher_sample_size}{Open} \\
Simulated power & \code{examples/paper/t\_test\_simulation.stat} & \studiolink{paper/t_test_simulation}{Open} \\
Interval precision & \code{examples/paper/response\_rate\_precision.stat} & \studiolink{paper/response_rate_precision}{Open} \\
Simon design guarantee & \code{examples/paper/simon\_design.stat} & \studiolink{paper/simon_design}{Open} \\
Simon design selection & \code{examples/paper/simon\_search.stat} & \studiolink{paper/simon_search}{Open} \\
COPD cited methods & \code{replication/copd/design\_cited.stat} & \studiolink{copd/design_cited}{Open} \\
COPD declared methods & \code{replication/copd/design\_declared.stat} & \studiolink{copd/design_declared}{Open} \\
COPD power & \code{replication/copd/power.stat} & \studiolink{copd/power}{Open} \\
\hline
\end{tabular}
\caption{Source files and Studio links for the article's demonstrations and
case studies.}
\label{tab:paper-studio}
\end{table}

\end{document}